\documentclass[adp,a4paper,fleqn]{w-art}

\usepackage{amssymb}
\usepackage{amsmath}
\usepackage{times,cite,w-thm}
\usepackage{graphicx}

\theoremstyle{plain}

\theoremstyle{definition}

\begin{document}
\DOIsuffix{theDOIsuffix}
\Volume{XX}
\Month{XX}
\Year{20XX}

\pagespan{1}{}

\Receiveddate{15 November 2009}
\Accepteddate{4 January 2010}
\Dateposted{ 21 February 2010}

\title{Dmitrij Volkov, super-Poincare group and \\Grassmann variables}
\author
{A.A. Zheltukhin \inst{1,2,3}%
\footnote{Corresponding author\quad E-mail:~\textsf{aaz@physto.se},
          Phone: 46\,8\,5537\,8743,
          Fax: 46\,8\,5537\,8601}}
\address[\inst{1}]{ Kharkov Institute of Physics and Technology, 
1, Akademicheskaya St., Kharkov, 61108, Ukraine }
\address[\inst{2}] {Fysikum, AlbaNova, Stockholm University, 
106 91, Stockholm, Sweden}
\address[\inst{3}] 
 {NORDITA,  
Roslagstullsbacken 23, 106 91 Stockholm, Sweden}

\maketitle

\begin{abstract}

A fundamental role of the Hermann Grassmann anticommuting variables both in
 physics and mathematics is discussed on the example of supersymmetry. 
The talk describes how the D. Volkov question about possibility of the existence 
of Nambu-Goldstone fermions, realized by the Grassmannian variables, resulted
 in the discovery of the super-Poincare group, its spontaneous breaking and gauging. 

\end{abstract}

\section{Introduction}

The famous Grassmannian variables discovered by Hermann G\"unther Grassmann 
in XIX century created the mathematical ground for the description of 
fermionic degrees of freedom in quantum field theory, and resulted 
in many outstanding discoveries in physics and mathematics.

Here I shall remind only one example demonstrating 
the fundamental role of the Grassmannian variables and their algebra 
in the discovery of supersymmetry.

\section{Spontaneous symmetry breaking and the Nambu-Goldstone bosons}

 The  phenomena of the spontaneous symmetry breaking was studied by Nambu \cite{Namb}, 
Goldstone \cite{Gold}, Bogolyubov \cite{Bog}, Schwinger \cite{Schw}, 
Weinberg \cite{Wein} and others in the 1960s. 
The massless and spinless 
Nambu-Goldstone particles associated with arbitrary spontaneously 
broken group $G$ were 
identified with  the coordinates of the coset space $G/H$ \cite{CCWZ, V1}
The internal symmetry group $G$ describes the symmetry of the Lagrangian of a physical 
system and 
its subgroup $H$ is the vacuum state symmetry. 
The approach of Volkov was based on the works 
by  Elie Cartan on symmetric 
spaces and exterior differential forms and resulted in the construction 
of $G$-invariant Phenomenological Lagrangians of the interacting 
N-G bosons \cite{V1}
\begin{equation}\label{L} 
\mathcal{L} = \frac{1}{2}Sp (G^{-1}dG)_{k}(G^{-1}dG)_{k}, \ \ \ G=KH,
\end{equation}
where the differential 1-forms $ G^{-1}dG=H^{-1}(K^{-1}dK)H + H^{-1}dH$ represent 
the vielbeins $ (G^{-1}dG)_{k}$ and the connection $(G^{-1}dG)_{h}$ associated with 
the symmetry subgroup $H$ transforming  one vacuum state to another. 
 Ferromagnet is a well-known example of the system with the spontaneously
 broken rotational $SO(3)$ symmetry with respect to spins in the Heisenberg Hamiltonian.
 Since for  ferromagnet  the symmetry of vacuum state is a subgroup $O(2)$
of  $SO(3)$ the N-G exitations turn out to be spin waves. 
The phenomenological Lagrangian of spin waves in (anti)ferromagnets, 
ferrits and in general case of spatially disordered media, taking into account 
new possible phases caused by {\it total} spontaneous breaking of the spin 
rotation  group, was constructed in \cite{VZB}.
At that time Volkov \cite{V2} put forward the
 question about the possibility of existence of the N-G fermions with 
spin $1/2$ and its study resulted in the super-Poincare 
group discovery \cite{VA}. 
The super-Poincare group was also independently discovered by Yu. Gol'fand, 
 E. Lichtman \cite{GL}, and J. Wess, B. Zumino \cite{WZ}. 
The basic motivation  of Gol'fand, Lichtman  was to construct 
quantum field theory with the parity violation. Wess, Zumino proposed  
generalization to four dimensions for the two-dimensional world-sheet supergauge
(superconformal) symmetry revealed by P. Ramond \cite{Ram},  A. Neveu, 
J. Schwartz \cite{NevSch}, J. Gervais, B. Sakita \cite{GerSak} in the 
models of the spinning strings and their world-sheet realization.
A crucial step in this way was the use of the anticommuting $c$-number
variables introduced by Hermann Grassmann. I shall talk about it on 
the basis of the original papers \cite{VA}.

\section{ The Lorentz and SL(2C) groups and their matrix realizations}

 The proper Lorentz group in $D=4$ Minkowski space is presented by 
real $4\times4$ matrices $\Lambda$ preserving both the scalar product of $4$-vectors
 $xy:=x_{m}y^{m}$ and the symmetric metric tensor $\eta_{mn}=diag(-1,1,1,1)$
\begin{equation} \label{vect}
\Lambda \eta  \Lambda^{T} =\eta,  \ \ det\Lambda=1, \ \  \eta^T=\eta.
\end{equation}
 The Lorentz group is locally isomorphic to the group $SL(2C)$ of complex $2\times2$
matrices $L$ preserving the scalar product $\psi\chi:=\psi_{\alpha}\chi^{\alpha}$  of the 
Weyl spinors $\psi_{\alpha},\chi^{\alpha}:=\varepsilon^{\alpha\beta}\chi_{\beta}$ and the 
antisymmetric metric 
tensor $\varepsilon_{\alpha\beta} \, (\varepsilon_{12}=\varepsilon^{21}=-1)$
\begin{equation} \label{spin}
L \varepsilon L^{T} =\varepsilon,  \ \  det L=1,  \ \ 
\varepsilon^T=-\varepsilon.
\end{equation}
The correspondence $\Lambda \rightarrow \pm L$ gives a two-valued representation of the 
proper Lorentz group and the $SL(2C)$ group has played essential role in the 
super-Poincare group discovery.
The Pauli matrices $\sigma_{i}$  together with the identity matrix  $\sigma_{0}$ form a 
basic set  $\sigma_{m}=(\sigma_{0}, \sigma_{i})$ in the space of $SL(2C)$ matrices.
The Lorentz covariant description demands the second  
set of the  Pauli matrices with the upper spinor indices 
 $\tilde\sigma_{m}:=( \tilde\sigma_{0}, \tilde\sigma_{i}):=(\sigma_{0}, -\sigma_{i})$ 
such that 
\begin{equation} \label{antc}
\{\sigma_{m}, \tilde\sigma_{n}\}= -2\eta _{mn},  \ \ \ 
Sp\sigma_{m}\tilde\sigma_{n}=-2\eta_{mn},
 \ \ \  \sigma^{m}_{\alpha\dot\alpha}\tilde\sigma_{m}^{\dot\beta\beta}
=-2\delta_{\alpha}^{\beta}\delta_{\dot\alpha}^{\dot\beta}.
\end{equation} 
The relativistic Pauli matrices $\sigma_{m}$ and $\tilde\sigma_{m}$ are Lorentz invariant 
analogously to the tensors $\eta_{mn}$ and $\varepsilon_{\alpha\beta}$.
Using two sets of the Pauli matrices yields the Lorentz covariant realization 
of the $\Lambda$ matrices in terms of the $L$ matrices 
\begin{equation}\label{map}
 \Lambda_{m}^{n}= -\frac{1}{2}Sp(\tilde\sigma_{m}L\sigma^{n}L^{+})
\end{equation} 
 showing the mentioned two-valuedness of the  $\Lambda \rightarrow  L$ mapping. 
The relation (\ref{map}) follows from the known correspondence between
the real Minkowski vectors $x^{m}$ and the Hermitian  matrices $X=X^{+}$ of 
 the $SL(2C)$ group
\begin{eqnarray} \label{herm}
X:=x^{m}\sigma_{m}=\left( \begin{array}{cc}
                       x^{0}+ x^{3}  & x^{1}-i x^{2}   \\
                        x^{1}+i x^{2}& x^{0}- x^{3}
                          \end{array} \right), \ \ \ x^{m}=-\frac{1}{2}Sp(\tilde\sigma^{m}X).
\end{eqnarray}
 Since  $detX=-x^{m}x_{m}$ the matrices $X$ realize   
 $SL(2C)$ transformations  
 \begin{equation} \label{rot}
X'=LXL^{+} \ \  \rightarrow \ \   det X'= -x^{'2}=detX=-x^{2},
\end{equation} 
which proves that the 4-vectors $x^{'m}$ and $x^{m}$ are connected 
by a Lorentz group transformation.  The representation  (\ref{map}) follows  from
Eq. (\ref{rot}) and the relation 
\begin{equation} \label{mapp}
X'=x^{'}_{m}\sigma^{m}= \Lambda_{m}^{n}x_{n}\sigma^{m}=Lx_{m}\sigma^{m}L^{+}.
\end{equation}
This information is all necessary to consider the Poincare group realization 
by $2\times2$ matrices.

\section{The triangle matrix realization of the Poincare group}

 The Hermitian matrices $X$ realize the transformations of the proper 
Poincare group $ x^{'}_{m}=  \Lambda_{m}^{n}x_{n} + t_{m}$  defined as follows
\begin{equation} \label{poinc}
X'=LXL^{+} + T,
\end{equation}
where $ T= t_{m}\sigma^{m}$. As a result,  any Poincare group 
element may be presented by the couple $(T,L)$ with their composition law given by
\begin{equation} 
\label{grol}
(T_{2}, L_{2})(T_{1}, L_{1})=(T_{3}, L_{3}):=(L_{2} T_{1} L_{2}^{+} + T_{2},  L_{2} L_{1}).
\end{equation}
It is well known that the group elements  $(T,L)$ are  presented by
complex $4\times4$  matrices  $G_{\wp}$                       
\begin{equation} \label{triang}
G_{\wp}=\left( \begin{array}{cc}
                       L & iTL^{+-1} \\
                       0 & L^{+-1}
                          \end{array} \right)=
\left( \begin{array}{cc}
                       1& iT\\
                       0 & 1
                          \end{array} \right)
\left( \begin{array}{cc}
                       L & 0\\
                       0 & L^{+-1}
                          \end{array} \right).
\end{equation}
This fact directly follows from the matrix multiplication
$$
\left( \begin{array}{cc}
                       L_{2} & iT_{2}L_{2}^{+-1} \\
                       0 & L_{2}^{+-1}
                          \end{array} \right)
\left( \begin{array}{cc}
                       L_{1} & iT_{1}L_{1}^{+-1} \\
                       0 & L_{1}^{+-1}
                          \end{array} \right)
=\left( \begin{array}{cc}
                       L_{2}  L_{1}& i(L_{2}T_{1}L_{2}^{+} + T_{2})( L_{2} L_{1})^{+-1} \\
                       0 & (L_{2} L_{1})^{+-1}
                          \end{array} \right).
$$
The factorization $G_{\wp}=K_{\wp}H$  (\ref{triang}) shows that the 
translation matrices  $T$  form a homogenious space under the Poincare group transformations
\begin{eqnarray}
G^{'}_{\wp} G_{\wp}=G_{\wp}^{''}=\left( \begin{array}{cc}
                       1& iT^{''}\\
                       0 & 1
                          \end{array} \right)
\left( \begin{array}{cc}
                       L^{''} & 0\\
                       0 & L^{''+-1}
                          \end{array} \right),  \nonumber   \\ 
L^{''}= L^{'} L, \ \ \    T^{''}= L^{'}T L^{'+} +  T^{'},  \label{grtr}
\end{eqnarray}
where $T^{'}=t_{m}^{'}\sigma^{m},
 \ \ T^{''}=t_{m}^{''}\sigma^{m}$,
which yield  the linear transformation $t_{m}^{''}=\Lambda_{m}^{n} t_{n} + t_{m}^{'}$
of the translation parameters $t_{m}$. 
It means that the parameters $t_{m}$ are identified with  the space-time 
coordinates $x_{m}$, respectively $T=X$ (\ref{herm}),
and give the represenation of the triangle matrices $G_{\wp}$ (\ref{triang}) 
in the form 
\begin{equation} \label{key}
G_{\wp}=\left( \begin{array}{cc}
                       L & iXL^{+-1} \\
                       0 & L^{+-1}
                          \end{array} \right)=
\left( \begin{array}{cc}
                       1& iX\\
                       0 & 1
                          \end{array} \right)
\left( \begin{array}{cc}
                       L & 0\\
                       0 & L^{+-1}
                          \end{array} \right)=K_{\wp}H.
\end{equation}
The presentation (\ref{key}) was one of the key elements Dmitrij  Volkov used
 in the construction of the super-Poincare group.

\section {The Grassmannian variables and the super-Poincare group}

The Volkov's idea was to separate the blocks of the $4\times4$  
matrix $K_{\wp}$  (\ref{key}) transforming it into the following  $5\times5$ triangle matrix
\begin{equation} \label{split}
K_{\wp}=\left( \begin{array}{cc}
                       1& iX \\
                       0 & 1
                          \end{array} \right) \ \ \rightarrow \ \ 
\left( \begin{array}{ccc}
                       1&0& iX\\
                       0&1&0 \\
                       0&0&1
                          \end{array} \right),
\end{equation}
and then to fill the empty places in the upper triangle block (\ref{split}) with 
 the anticommuting Grassmannian Weyl spinors $\theta_{\alpha}$, because they were supposed 
to play the role of N-G particles with spin $1/2$ subjected to the Fermi 
statistics $\theta_{1} \theta_{2}=-\theta_{2} \theta_{1}$, and consequently 
  $(\theta_{1})^2=(\theta_{2})^2\equiv0$,
\begin{equation} \label{supnkr}
K_{S\wp}=\left( \begin{array}{ccc}
                        1&\theta& iZ\\
                        0&1&\theta^{+} \\
                        0&0&1
                      \end{array} \right ), \ \ \ \theta =\left(\begin{array}{c} \theta_{1} \\
 \theta_{2} \end{array} \right),  \ \ \  
\theta^{+}=( \bar\theta_{\dot 1}, \bar\theta_{\dot 2}),
\end{equation}
where the complex matrix $Z$, defined  by the nilpotent shift of $X$
\begin{equation} \label{Zmatr}
Z= X -\frac{i}{2}\theta\theta^{+} \ \ \rightarrow \ \ 
z_{\alpha\dot\beta}=x_{\alpha\dot\beta} - \frac{i}{2}\theta_{\alpha}\bar\theta_{\dot\beta},
\end{equation}
 was substituted instead of $X$. 
Such a nontrival complexification of $X$ has been dictated by the condition 
to preserve the composition law
\begin{equation} \label{grplaw}
G'_{S\wp}G_{S\wp}=G''_{S\wp}
\end{equation}
for the  matrices $G_{S\wp}$
\begin{equation} \label{gpnkr}
G_{S\wp}=\left( \begin{array}{ccc}
                        1&\theta& iX +\frac{1}{2}\theta\theta^{+} \\
                        0&1&\theta^{+} \\
                        0&0&1 \end{array} \right )
\left( \begin{array}{ccc}
                        L&0& 0 \\
                        0&1& 0 \\
                        0&0& L^{+-1} \end{array} \right)
\end{equation}
  extending the Poincare group matrices $G_{\wp}$ (\ref{triang}). 
The substitution of $G_{S\wp}$ together with $ G'_{S\wp}$ and  $ G''_{S\wp}$, 
presented in the form similar to (\ref{gpnkr}), in Eq. (\ref{grplaw}) 
has revealed the transformation law \cite{VA}
\begin{equation} \label{supoilaw}
L^{''}= L^{'} L, \ \ \   \theta^{''}= L^{'}\theta + \theta',  \ \ \
X^{''}= L^{'}X L^{'+} + X^{'} + \frac{i}{2}(L^{'}\theta\theta^{'+} 
- \theta^{'}\theta^{+} L^{'+})
\end{equation} 
 generalizing the Poincare group law (\ref{grtr}).
 The corresponding transformations of the complex $Z$-matrices (\ref{Zmatr}) 
have the form:
$
Z^{''}= L^{'}Z L^{'+} -i \theta^{'}\theta^{+}L^{'+}
$. 

The relations (\ref{supoilaw}) are the transformations of the required 
super-Poincare group. They demonstrate a fundamental role of the Grassmannian 
variables for the discovery of the super-Poincare group.

\section {Supersymmetry and superalgebra}

 The supersymmetry transformations in the component form 
\begin{equation} \label{susy}
\theta'_{\alpha}= \theta_{\alpha} + \xi_{\alpha}, \ \ 
\bar\theta'_{\dot\alpha}= \bar\theta_{\dot\alpha} + \bar\xi_{\dot\alpha}, \ \
x^{'}_{\alpha\dot\alpha}=x_{\alpha\dot\alpha} + 
\frac{i}{2}( \theta_{\alpha}\bar\xi_{\dot\alpha}- \xi_{\alpha}\bar\theta_{\dot\alpha} )
\end{equation} 
are extracted from the matrix representation (\ref{supoilaw}) 
by choosing $ L=I_{2\times2},  X^{'}=0$ and $\theta'\equiv\xi$, 
where $I_{2\times2}$ is the $2\times2$ identity matrix. 
The supersymmetry generators $Q^{\alpha}$ and their complex 
conjugate  $\bar Q^{\dot\alpha}:= -(Q^{\alpha})^{*}$ 
\begin{equation} \label{gener}
Q^{\alpha}=\frac{\partial}{\partial\theta_{\alpha}} - \frac{i}{2} \bar\theta_{\dot\alpha}
\frac{\partial}{\partial x_{\alpha\dot\alpha}}, \ \ \
\bar Q^{\dot\alpha}= \frac{\partial}{\partial\bar\theta_{\dot\alpha}} 
- \frac{i}{2} \theta_{\alpha}
\frac{\partial}{\partial x_{\alpha\dot\alpha}}
\end{equation} 
 together with the translation generator $P^{m}=i\frac{\partial}{\partial x_{m}}$ 
form the superalgebra
\begin{eqnarray}\label{susyalg}
\{ Q^{\alpha}, \bar Q^{\dot\alpha} \}=-i\frac{\partial}{\partial x_{\alpha\dot\alpha}}
=\frac{1}{2}\tilde\sigma_{m}^{\dot\alpha\alpha}P^{m}, \\
\{ Q^{\alpha},  Q^{\beta} \}=\{ \bar Q^{\dot\alpha}, \bar Q^{\dot\beta} \}
=[ Q^{\alpha},P^{m}]=[ \bar Q^{\dot\alpha},P^{m}]=0 \nonumber
\end{eqnarray}
which is the supersymmetry algebra. 
The supersymmetry transformations (\ref{susy}) together  with 
 their non-zero anticommutator are presented in the equivalent Dirac bispinor form 
after transition to the Majorana spinors
\begin{eqnarray}\label{gamsusy}
\delta\theta=\xi, \ \ \  \  \delta\bar\theta=\bar\xi,  \ \ \ \
\delta x_{m}=-\frac{i}{4}(\bar\xi\gamma_{m}\theta),  \ \ \ \
\{ Q_{a}, Q_{b}\}= \frac{1}{2}(\gamma_{m}C^{-1})_{ab}P^{m},
\end{eqnarray}
where  $\bar\theta=\theta^{T}C$ with the antisymmetric matrix of the charge conjugation 
$
C^{ab}=  \left( \begin{array}{cc}
                       \varepsilon^{\alpha\beta}&0\\
                       0 & \varepsilon_{\dot\alpha\dot\beta}
                          \end{array} \right)$ and 
$ 
Q_{a}= \frac{\partial}{\partial\bar\theta^{a}} - \frac{i}{4}
 (\gamma_{m}\theta)_{a}\frac{\partial}{\partial x_{m}}.
$
The Majorana spinors and the 
 Dirac $\gamma$-matrices in (\ref{gamsusy}) are defined as follows
 \begin{eqnarray}\label{gamma}
\theta _{a}=\left(\begin{array}{c} \theta_{\alpha} \\ 
 \bar\theta^{\dot\alpha} \end{array} \right), \ \ 
\xi_{a}=\left(\begin{array}{c} \xi_{\alpha} \\ 
 \bar\xi^{\dot\alpha} \end{array} \right),  \ \  
\gamma_{m}=  \left( \begin{array}{cc}
                       0 & \sigma_{m}\\
                       \tilde\sigma_{m} & 0
                          \end{array} \right),  \ \ 
\{\gamma_{m}, \gamma_{n}\}= -2\eta _{mn}.
\end{eqnarray}
The transformations (\ref{susy}) or  (\ref{gamsusy}) are the 
transformations of the $N=1$ supersymmetry. 
The integer number $N$ denotes the number of 
the Weyl spinors $\theta_{\alpha}$ accompanying the coordinates $x_{m}$. 
The case  $N=1$ realizes a simple case of two complex 
 Grassmannian variables $\theta_{1}$ and  $\theta_{2}$. 
Actually Volkov applied his idea in 
more general form considering  the case of arbitrary $N$.

\section{Unification of the Poincare group and internal symmetries}

 In general form the Volkov's idea was to split the blocks of the $4\times4$  
matrix $K_{\wp}$ in the representation (\ref{key}) replacing it 
 by the $(4+N)\times (4+N)$ 
triangle matrix $K^{(ext)}_{S\wp}$ including the Grassmannian Weyl 
spinors $\theta^{I}_{\alpha}$ which had the index $I=1,2,...,N$ of an 
internal symmetry group (e.g. $SU(N)$). The new index enumerated the 
columns of the $2 \times N$ restangular $\theta$-submatrix. 
Simultaneously the Lorentz matrix $H$ (\ref{key}) was 
extended to the $(4+N) \times (4+N)$ matrix by the 
 addition of the $N \times N$  block submatrix $U_{N \times N}$ of 
the internal symmetry. The procedure resulted in the factorizable triangle matrix 
\begin{equation} \label{extpoinc}
G^{(ext)}_{S\wp}=K^{(ext)}_{S\wp}H^{(ext)}=
\left( \begin{array}{ccc}
                        1&\theta& iX +\frac{1}{2}\theta\theta^{+} \\
                        0&I_{N \times N}&\theta^{+} \\
                        0&0&1 \end{array} \right )
\left( \begin{array}{ccc}
                        L&0& 0 \\
                        0&U_{N \times N}&\theta^{+}\\
                        0&0& L^{+-1} \end{array} \right)
\end{equation}
 associated with the $N$-extended super-Poincare group and its transformations 
are  given by the  following matrix relations
\begin{eqnarray} \label{extsupoilaw}
L^{''}= L^{'} L, \ \ \ U^{''}= U^{'}U,  \ \ \   \theta^{''}= L^{'}\theta  U'^{-1}
+ \theta',  \\
X^{''}= L^{'}X L^{'+} + X^{'} + \frac{i}{2}(L^{'}\theta U'^{-1}\theta^{'+} 
- \theta^{'} U'\theta^{+} L^{'+}), \nonumber
\end{eqnarray}  
where $U\equiv U_{N \times N}$. The law (\ref{extsupoilaw}) was derived
 by the  substitution of $G^{(ext)}_{S\wp}$ 
(\ref{extpoinc}) for  $G_{S\wp}$ (\ref{gpnkr}) in the composition law (\ref{grplaw}).
The $N$-extended supersymmetry transformations, encoded in the matrix representation
 (\ref{extsupoilaw}),  generalize the  $N=1$ supersymmetry 
transformations (\ref{susy})
\begin{equation} \label{extsusy}
\theta'^{I}_{\alpha}= \theta^{I}_{\alpha} + \xi^{I}_{\alpha}, \ \ 
\bar\theta'_{\dot\alpha I}= \bar\theta_{\dot\alpha I} + \bar\xi_{\dot\alpha a}, \ \
x^{'}_{\alpha\dot\alpha}=x_{\alpha\dot\alpha} + \frac{i}{2}
(\theta^{I}_{\alpha}\bar\xi_{\dot\alpha I} -\xi^{I}_{\alpha}\bar\theta_{\dot\alpha a})
\end{equation} 
for the extended superspace $(x^{m},\theta^{I}_{\alpha},\bar\theta_{\dot\alpha I})$. 
The extension of the Minkowski space to the superspace has revealed the way to bypass the 
 Coleman-Mandula no-go theorem for the unification of the internal and space-time symmetries.

\section{The Nambu-Goldstone fermions with spin $1/2$}

The next important step made by  Volkov was the introduction of the supersymmetry invariant 
differential forms generalizing the famous Cartan $\omega$-forms for the case 
of space with the  Grassmannian coordinates. For the above considered coset space
$G^{(ext)}_{S\wp}/H^{(ext)}$ (\ref{extpoinc}), realized by the matrices $K^{(ext)}_{S\wp}$, 
the corresponding  $\omega$-forms appear  as the matrix blocks in the product 
\begin{eqnarray} \label{wmatr}
{K^{(ext)}_{S\wp}}^{-1}dK^{(ext)}_{S\wp}=
\left( \begin{array}{ccc}
                        1&-\theta& (iZ)^{+} \\
                        0&I_{N \times N}&-\theta^{+} \\
                        0&0&1 \end{array} \right )
\left( \begin{array}{ccc}
                        0&d\theta& idZ \\
                        0&0_{N \times N}&d\theta^{+}\\  \nonumber
                        0&0&0 \end{array} \right) \\
=
\left( \begin{array}{ccc}
                        0&d\theta&idX +\frac{1}{2}(d\theta\theta^{+} - \theta d\theta^{+}\\
                        0&0_{N \times N}&d\theta^{+} \\ 
                        0&0&0 \end{array} \right )
\end{eqnarray}  
and in the component form they are given by the following expressions 
\begin{eqnarray} \label{wforms}
\omega_{\alpha}^{I}= d\theta_{\alpha}^{I}  ,  \ \ \    
\bar\omega_{\dot\alpha I}=d\bar\theta_{\dot\alpha I}   ,  \ \ \  
\omega_{\alpha\dot\alpha}=dx_{\alpha\dot\alpha}-
\frac{i}{2}( d\theta_{\alpha}^{I} \bar\theta_{\dot\alpha I} - 
\theta_{\alpha}^{I} d\bar\theta_{\dot\alpha I}). 
\end{eqnarray}  
In the bispinor representation the fermionic and bosonic one-forms (\ref{wforms})  are 
\begin{eqnarray} \label{biwforms}
\omega= d\theta ,  \ \ \    
\bar\omega=d\bar\theta_  ,  \ \ \  
\omega_{m}=dx_{m} -
\frac{i}{4}( d\bar\theta\gamma_{m}\theta ).
\end{eqnarray} 
The $\omega$-forms are the building blocks for the construction of supersymmetric 
 actions of the interacting N-G particles. 
To construct the invariant actions Volkov generalized the Cartan method of the exterior 
differential forms for the case of superspace and found invariant hyper-volume, 
imbedded in the superspace, and other invariants. 
Since the invariant action of the N-G fermions has to include the factorized volume element 
$d^{4}x$ it  strongly restricts the 
structure of the admissible combinations of the $\omega$-forms. If 
 the combination is given by a product of the $\omega$-forms (\ref{wforms})
and their differentials, it should have the general number of the differentials equal to four. 
The conditition is satisfied by the well known invariant  \cite{VA}
\begin{eqnarray} \label{volum}
 d^{4}V=\frac{1}{4!}\varepsilon_{mnpq}\omega^{m}\wedge\omega^{n}\wedge\omega^{p}\wedge\omega^{q}, 
\end{eqnarray} 
where the symbol $\wedge$ sets the external product,
 that gives  the natural supersymmetric extension of the volume element $d^{4}x$ 
of the Minkowski space.
The supersymmetric volume (\ref{volum}), invariant also under the Lorentz
 and unitary groups, does not contain the spinorial 
one-forms $\omega_{\alpha}^{I}$ and $\bar\omega_{\dot\alpha I}$, but they appear, 
 e.g. in  the following invariant products  
\begin{eqnarray}\label{volhor}
 \Omega^{(4)}=\omega_{\alpha}^{I}\wedge\bar\omega_{\dot\beta I}\wedge 
\tilde\sigma^{\dot\beta\alpha}_{m} d\wedge\omega^{m},\ \ \ 
 {\tilde\Omega}^{(4)}=\varepsilon^{\alpha\beta}\omega_{\alpha}^{I}\wedge
\omega_{\beta}^{J}\wedge\bar\omega_{\dot\alpha I}\wedge 
\bar\omega_{\dot\beta J}\varepsilon^{\dot\alpha\dot\beta},
\end{eqnarray} 
where $d\wedge\omega^{m}$ denotes the external differential of $\omega^{m}$ \cite{VA}.
 The Volkov's idea for the construction of the  N-G fermion action
 was to use the pullback of the differential form $d^{4}V$ (\ref{volum}), 
and its generalizations similar to (\ref{volhor}), on 
 the $4$-dimensional Minkowski subspace of the superspace. The pullback is 
 realized by the parametrization of $\theta$ by the Minkowski coordinates $x_{m}$.
 As a result, the differential forms  
$\omega_{m}$ (\ref{biwforms}) and  $d^{4}V$ (\ref{volum}), e.g. for the case $N=1$, 
take the form
\begin{eqnarray}\label{pullb}
\omega_{m}= 
(\delta_{m}^{n} -\frac{i}{4}\frac{\partial\bar\theta}{\partial x_{n}}\gamma_{m}\theta)
dx_{n}=W_{m}^{n}dx_{n}, \  \  \  
d^{4}V=\det W d^{4}x.
\end{eqnarray}  
Due to the spinor $\theta$ dependence on $x$,  it was identified with 
the N-G fermionic field $\psi(x) =a^{-1/2}\theta(x)$ with spin $1/2$, where $a$ 
is the coupling constant $[a]=L^{4}$.
The constant adds the correct dimension $L^{-3/2}$ to the 
fermionic field $\psi(x)$  and its substitution for $\theta$ in (\ref{pullb})
gives the Volkov-Akulov action \cite{VA}
\begin{eqnarray} \label{action}
S=\frac{1}{a}\int\det W d^{4}x. 
\end{eqnarray}  
 An explicit form of the action $S$ (\ref{action}) with $W=W(\psi,\partial_{m}\psi)$ is 
\begin{eqnarray} \label{action'}
S= \int d^{4}x [\, \frac{1}{a} + T_{m}^{m} + \frac{a}{2}(T_{m}^{m}T_{n}^{n}-
T_{m}^{n}T_{n}^ {m}) + a^2 T^{(3)}+ a^3 T^{(4)}  \,],
\end{eqnarray} 
where $T^{(3)}$ and   $T^{(4)}$ code the interaction terms of the N-G fermions that are 
cubic and quartic in the particle momentum $\partial^{m}\psi$,
 and $T_{m}^{n}$ is defined by the following relations 
\begin{eqnarray} \label{acterm}
W_{m}^{n}= \delta_{m}^{n} +aT_{m}^{n},\ \ \  T_{m}^{n}= 
-\frac{i}{4}\partial^{n}\bar\psi\gamma_{m}\psi.
\end{eqnarray} 
The first term in (\ref{action'}), unessential for the Minkowski space,
 is interesting because of its possible connection with the cosmological term 
in a curved superspace.
The second term coincides with the free Dirac action of the massless fermion 
field $\psi(x)$ and  has the form
\begin{eqnarray} \label{dirac}
S_{0}=\int d^{4}x T_{m}^{m}
= -\frac{i}{4}\int d^{4}x\partial^{m}\bar\psi\gamma_{m}\psi,
\end{eqnarray} 
proving that the  N-G field $\psi(x)$ actually carries spin $1/2$. 
The cubic $ T^{(3)}$ and quartic $ T^{(4)}$ terms in the particle momenta have the 
following structure \cite{VA}
\begin{eqnarray} \label{vertex}
 T^{(3)}= \frac{1}{3!}\sum_{p} (-)^{p} T_{m}^{m} T_{n}^{n} T_{l}^{l},\ \ \ \  \
T^{(4)}= \frac{1}{4!}\sum_{p} (-)^{p} T_{m}^{m} T_{n}^{n} T_{l}^{l}T_{k}^{k},
\end{eqnarray} 
where the sum  $\sum_{p} $ corresponds to the sum in all permutations of the 
subindices in the products of the tensors $ T_{m}^{m}$,
 and these terms describe the vertexes with six and eight N-G fermions respectively.
The  vertexes (\ref{vertex}) were analyzed in \cite{Kuz} and the vanishing 
of  the  quartic term $T^{(4)}$ was observed there
\footnote{Sergei Kuzenko kindly informed me  about Ref.\cite{Kuz}.}. 

The Volkov's  approach clearly shows how to construct the higher degree terms 
in the derivatives of the N-G fields to get possible 
supersymmetric generalizations of the Volkov-Akulov action.
In general case the combinations of the $\omega$-forms (\ref{wforms}), 
admissible for the higher order invariant action, are the homogenious 
functions of the degree four with respect of the differentias $dx$ and $d\psi$. 
The latter condition guarantees the factorization of the volume element $d^{4}x$ 
in the generalized action integral. To restrict the number of such type invariants 
 Volkov proposed to use the minimality 
condition with respect to the degree of the derivatives $\partial\psi/\partial x$
in the general nonlinear action for  the N-G fermions
\begin{equation}\label{genract}
S=\int d^{4}x  L(\psi, \partial\psi/\partial x)
\end{equation} 
which corresponds to taking into account only the lowest degrees of the momenta of N-G fermions 
in their scattering matrix. 
To find the degree of $\partial\psi/\partial x$ in different
invariants it was observed that such derivatives appear from the differentials $d\psi$ 
in the fundamental one-forms (\ref{wforms}). In addition, the spinor one-forms create
one derivative $\partial\psi/\partial x$,  but the vector form terms either do not contain 
the $\psi$ fields or contain one derivative  $\partial\psi/\partial x$ accompanied by $\psi$. 
As a result,
the number of the derivatives $\partial\psi/\partial x$  with respect to the whole number of
 the fields is lower in the vector differential one-form than in the spinor ones. 
Also, the invariants
including the differential of the $\omega$-forms, like ${\tilde\Omega}^{(4)}$ 
in (\ref{volhor}), have the higher degree in $\partial\psi/\partial x$
in comparison with the $\omega$-forms themselves. Thus, the demand of the minimality of 
the degree of derivatives in $S$ (\ref{genract}) will be 
satisfied if the admisible invariants contain only the vector differential one-forms $\omega_{m}$. 
 
Moreover, Volkov has developed the general method for the supersymmetric  
inclusion of the N-G particle interactions with other fields. 
 For a given field $\Phi$, carrying the spinor and unitary indices, 
its differential  $d\Phi$ has to  be used on the same level as the 
supersymmetric $\omega$-forms (\ref{wforms}) in the application of  the above 
described procedure.  
  The only restriction on the admissible terms including  $\Phi$ is the demand 
of their invariance
 under the Lorentz and the unitary groups.  The invariant interaction of the  
massive Dirac field with 
the N-G fermions was considered in \cite{VA} as an instructive example of the 
described procedure.

The interest to the problem of the spontaneous 
symmetry breaking was recently strengthened in connection with the 
paper \cite{Seib} (and Refs. there), 
where an approach to this fundamenental problem, based on new physical observations and 
 superfield effective Lagrangian for the  N-G fermions, 
 has been discussed \footnote{ Paolo Di Vecchia attracted my attention to Ref. \cite{Seib}}.
One can hope that the Volkov's geometrical approach will strongly help in the approach development.

\section{Gauging the super-Poincare group and supergravity}
 
Taking into account spontaneously broken character of the super-Poincare group,
 Volkov had immediately understood that its gauging would create
 the massive Rarita-Schwinger gauge field (gravitino) of spin $3/2$ 
absorbing the N-G fermion because 
of the Higgs effect. Realization of this idea led Volkov to the discovery of supergravity -  
 the new physical theory - where graviton gets the gravitino as a superpartner. 
 The corresponding supergravity action was published 
by D. Volkov and V. Soroka in 1973 \cite{VS}.
 Next papers in supergravity were published in 1976 by S. Deser, B. Zumino \cite{DZ}, 
S. Ferrara, D. Friedmann, P. Van Niewenhuizen \cite{FFN} and 
towards the end of the 1970's SUSY and SUGRA  became the generally accepted approaches in
 the field theory. The exciting story of this discovery was presented 
by Volkov in his talk in Erice \cite{Verice}. 
 The maximally extended $N=8$ supergravity and supersymmetry are discussed by H. Nicolai
in this volume \cite{Nic}.   
 
\section{From "Ausdehnungslehre" by Hermann Grassmann to superstrings}

The Volkov's method of the construction of the action integrals for the N-G fermions 
as supersymmetric hypervolumes (and their generalizations) 
imbedded in superspace has shown the general way of 
constructing the action integrals for superparticles and for extended supersymmetric objects 
superstrings, super $p$-branes \cite{GSW},  \cite{Pol}. 
Actually, in this way  $d^{4}V$ 
(\ref{volum}) may be interpreted as the world-volume of a super $3$-brane imbedded 
in the superspace, if the super coordinates $(x, \theta$)  are
parametrized by the  world-volume 
coordinates $(\tau,\sigma^1, \sigma^2, \sigma^3)$ of the super $3$-brane. 
All that refers us to the revolutionary book 
"Ausdehnungslehre" by Hermann Grassmann \cite{Gras} with his theory of extensions and the  
algebra of exterior forms that so inspired William Clifford and Elie Cartan. 
Due to the clear explanations by Felix Klein \cite{Klein} about the Grassmann idea to 
treat the finite pieces of lines, planes and hypersurfaces as pure
 geometric objects out of coordinates and metric properties, one can see that
 Grassmann had anticipated the role of 
superstrings and super p-branes that so naturally 
 unified the Grassmannian variables with his theory of extensions.
 On behalf of mathematics the Grassmann ideas 
inspired Felix Berezin and others to create the supermathematics \cite{Berez}. 
The limited space will not allow me to remind about many outstanding physicists 
and mathematicians and their great contributions to the development of supersymmetry, 
supergravity and supermathematics.

Thus, the brilliant algebraic, geometric and physical ideas 
of Hermann G\"unther Grassmann ran way ahead
 of their time and have created the basic  mathematical and physical
 entities which became the foundation   
for the construction of space-time theory, the modern quantum field theory, 
supersymmetry and string theory.

\noindent{\bf Acknowledgments}

I would like to thank Prof. Mariusz Dabrowski for inviting me to participate
at the "Grassmannian Conference in Fundamental Cosmology", 
valuable collaboration and support. 
I would also like to thank John Barrow, Paolo Di Vecchia,  
Anthony Lasenby, Alexej Starobinsky, Larus Thorlacius for interesting discussions, 
and Sergei Kuzenko, Vladimir Tkach for their kind letters.
I am grateful to the University of Szczecin, Fysikum at Stockholm University 
and Nordic Institute for Theoretical Physics Nordita for kind hospitality. 
This research was supported in part by Nordita.


\begin{thebibliography}{99}

\bibitem{Namb}
Y. Nambu,  Phys. Rev.  {\bf 117}, 648 (1960);
Phys. Rev. Lett. {\bf 4}, 380  (1960); \\
Y. Nambu and G. Jona-Lasinio  Phys. Rev. {\bf 122}, 345 (1961); ibid. {\bf 124}, 246 (1961).
\bibitem{Gold}
J. Goldstone, Nuov. Cim. {\bf 19}, 155  (1961).
\bibitem{Bog}
N. N. Bogolyubov, Lectures on Quantum Statistics, London (1970);  
Selected  Works, Vol.3 (in Russian)(Naukova Dumka, Kiev, 1971) .
\bibitem{Schw}
J. Schwinger, Phys. Lett. {\bf B24}, 473 (1967).
\bibitem{Wein}
S.Weinberg,  Phys. Rev. {\bf 166},  1568  (1968).
\bibitem{CCWZ}
S. Coleman, J. Wess and B. Zumino, Phys. Rev. {\bf 177},  2239 (1969); \\
C. Callan, S. Coleman, J. Wess and B. Zumino, Phys. Rev. {\bf 177}, 2247 (1969).
\bibitem{V1}
D.V. Volkov,
Phenomenological Lagrangian of the Goldstone particle interactions,
 Preprint ITF-69-75, Kiev, (1969) (in Russian); 
Sov. J. Part. Nuclei  {\bf 4}(1), 3  (1973).
\bibitem{VZB}
D.V. Volkov, A.A. Zheltukhin and Yu.P. Bliokh,
Phys. Solid State  {\bf 13}, 1396   (1971);\\
D.V. Volkov and  A.A. Zheltukhin,
Sov. Phys.- JETP {\bf 51},  937 (1980).
\bibitem{V2}
D.V. Volkov,
 Phenomenological Lagrangians invariant under symmetry groups including 
the Poincare group as a subgroup,
  Preprint N114, (in Russian)(Lebedev Phys.Inst., Moscow,1971).
\bibitem{VA}
D.V. Volkov and V.P. Akulov, JETP Letters {\bf 16}, 478 (1972);
 Phys. Lett. {\bf B46}, 109  (1973);  Theor. Math. Phys.  {\bf 18}, 28 (1974).
\bibitem{GL}
Yu. Gol'fand and E. Lichtman, JETP Letters  {\bf 13}, 323 (1971).
\bibitem{WZ}
J. Wess and B. Zumino, Nucl. Phys.  {\bf B70},  39  (1974).
\bibitem{Ram}
P. Ramond,  Phys. Rev. {\bf D3},  2415 (1971).
\bibitem{NevSch}
A. Neveu and J.H. Schwartz,  Nucl. Phys. {\bf B31}, 86 (1971).
\bibitem{GerSak}
J.L. Gervais and B. Sakita,
  Nucl. Phys.  {\bf B34},  832 (1971).
\bibitem{Kuz}
S.M. Kuzenko and S.A. McCarthy, JHEP {\bf 05} 012  (2005).
\bibitem{Seib}
Z. Komargordski and N. Seiberg,
From Linear SUSY to Constrained  Superfields, arXiv:0907.2441v3 [hep-th].
\bibitem{VS}
D.V. Volkov and V.A. Soroka, JETP Letters {\bf 18}, 312  (1973);
 Theor. Math. Phys. {\bf 20},  829 (1974).
 \bibitem{DZ}
S. Deser and B. Zumino, Phys. Lett. {\bf 62}, 335 (1976).
 \bibitem{FFN}
S. Ferrara, D. Friedmann and P. Van Niewenhuizen, 
Phys. Rev.  {\bf D13}, 3214 (1976).
\bibitem{Verice}
D.V. Volkov, Supergravity before 1976, Proceedings of International 
Conference on History of  Original Ideas  and  Basic Discoveries in Particle Physics,
 edited by  H.B. Newman and T. Ypsilantis (Plenum Press, New York 1996), pp. 663-675.
\bibitem{Nic}
H. Nicolai, 
Ann. Phys.(Berlin) {\bf 3-4} (2010), to be published.
\bibitem{Gras}
Hermann G\"unther Grassmann,
Die Ausdehnungslehre, Vollst\"andig und in strenger Form bearbeitet",
(T.C.F. Enslin, Berlin, 1862).
\bibitem{Klein}
Felix Klein,
Vorlesungen \"uber die Entwicklung der Mathematik  im 19. Jahrhundert, Teil I (Springer, Berlin, 1926).
\bibitem{Berez}
F.A. Berezin,
Introduction to Algebra and Analysis with Anticommuting Variables 
(Moscow State University Press, Moscow, 1983).
\bibitem{GSW}
M.B. Green, J.H. Schwartz and E. Witten, 
Superstring Theory (Cambridge University Press, Cambridge, 1987).
\bibitem{Pol}
J. Polchinski,
String Theory Vol. II: Superstring Theory and Beoynd (Cambridge University Press, Cambridge, 1998).


\end{thebibliography}
\end{document}